\documentstyle[11pt,newpasp,twoside,epsf]{article}
\begin{document}

\title{High Speed Interconnects and Parallel Software Libraries:\\
       Enabling Technologies for the NVO\footnotemark[1]}
\author{Jeremy Kepner and Janice McMahon}
\affil{MIT Lincoln Laboratory, Lexington, MA}

\footnotetext[1]{This work is sponsored by DARPA, under Air Force Contract
F19628-00-C-0002.  Opinions, interpretations, conclusions and
recommendations are those of the author and are not necessarily endorsed
by the United States Air Force.}

%\begin{abstract}
%\end{abstract}

\section{Introduction}
  The National Virtual Observatory (NVO) will directly or indirectly
touch upon all steps in the process of transforming raw observational
data into ``meaningful'' results.  These steps include:
\begin{itemize}
  \item[(1)] Acquisition and storage of raw data.
  \item[(2)] Data reduction (i.e. translating raw data into source detections).
  \item[(3)] Acquisition and storage of detected sources.
  \item[(4)] Multi-sensor/multi-temporal data mining of the
             products of steps (1), (2) and (3).
\end{itemize}
The highly distributed nature of the NVO places new twists on all of
these steps.  Future NVO research is likely to focus on developing the
software tools necessary for Step (4) as well as the methods for
``federating'' data from Steps (1) and (3).  However, past experience
with individual surveys indicates that Step (2) has dominated computer
software and hardware costs and may have a large impact on the NVO.

  Federation of data sets from multiple institutions, which is a primary
NVO goal, will be made significantly easier if improvement of the data
reduction pipeline software is also undertaken. Addressing the
challenges of Step (2) for the NVO can be accomplished by significantly
improving the software environment for data reduction pipelines. 
Although the NVO can and should influence this effort it may be outside
of the NVOs core activities.

  The rest of this paper presents a further analysis of the computing
and networking requirements of the NVO and provides a discussion of
some of the challenges and solutions for addressing data reduction for
these massive NVO data sets.

\section{Large Survey Requirements}
  Large area imaging surveys are generating data at an exponentially
increasing rate.  Reducing this data is a significant hardware and
software challenge that is likely to push the limits of computing for
some time to come.  The computing requirements of these surveys are best
understood in terms of the number of operations required to process a
single pixel or a single detection.  By looking at the data reduction
pipeline in these terms it is easy to determine the overall computing
needs of a given data stream by multiplying the data rate by the per
pixel processing requirement.

\subsection{Computing Requirements}
  Figure 1 shows the steps that are common to most data reduction
pipelines.  The primary driver is the matched filtering step which
involves 2D convolutions of the entire image.  Such convolutions can
easily result in as many as $10^4$ operations per pixel and requires a
sizable amount of computing to keep up with the data rate.  Detection
processing also involves 2D convolutions but of a sparser nature. 
Although the real-time processing requirements of a survey may be
readily satisfied by a few tens of computers, it is quite common to
re-run the data processing pipeline on several years worth of data after
the pipeline software has been upgraded.  In such cases, hundreds or
even thousands of computers will be needed to re-process the data in a
reasonable time.

  For example, to support a camera with a real time data
rate of 100 million pixels per second (e.g. the LSST) requires a
computer system that can {\it deliver} 100 Gigaflops, which is
approximately 1000 of todays state of the art workstations.  To
re-process such a data stream could require millions of such computers.

\subsection{Software Requirements}
  The primary difficulty in developing high performance pipeline
software is the large number of systems architecture issues (number of
processors, total memory, network bandwidth, disk bandwidth, ...) that
need to be considered in order to keep up with the data rate (see
Figure 2). In other words, the software pipeline becomes highly tuned
to the system, which in turn increases the size of the code and the
expertise necessary to maintain the code.  In addition, upgrading to a
new system can require a significant re-write.  All of these issues
mean that there is very little code re-use across different
data pipelines.

  Unfortunately, the need for highly tuned software will only grow in
the future as next generation computers incorporate more complex
features (e.g., multi-processor chips, on chip vector units,
multi-threading, ...). While it is now possible to get 1\% or 10\% of a
computers peak performance with little or no optimization, in the
future an un-optimized program can expected to give less than 0.1\% of
peak performance. In other words, the price of not being optimized to
the hardware will increase.

\section{Enabling Technologies}
  The primary goals of a computer system architect are to manage
complexity and to create systems that avoid the need for heroic
programming efforts to meet the program milestones.  In the context
of a real-time data reduction pipeline this means developing
simplified abstractions of the hardware and software so as to isolate
the specifics of the application from the specifics of the hardware.

  In the previous section, the argument for complex, highly parallel
computing systems was presented.  Without such systems it
will require months to analyze a single nights worth of observations,
and significantly longer to re-process a significant portion of a survey.

\subsection{High Speed Interconnects}
  One way to create a less complex parallel computing systems is to
invest in extremely capable networks.  Minimizing network usage is the
single most common optimization step performed in writing parallel
software. High performance networks greatly alleviate the pressure on
the programmer to implement to the specific hardware.  In other words,
it is incumbent the system architect to assemble a ``balanced'' system
(i.e. computation vs. communication). Fortunately, the rapid rise of the
Internet and cluster computing has driven the need for ever more capable
interconnects.  As a result a variety of technologies will be available
for producing systems that are well balanced for data processing
pipelines (see Figure 3).

\subsection{High Performance Parallel Software Libraries}
  A fast method for implementing high performance data pipelines is to
re-use already optimized code.  The best way is to leverage existing
libraries (e.g. Lapack, ScaLapack, FFTW, VSIPL, ...) developed by other
communities (see Figure 4).  These software packages remove the majority
of the effort required to achieve optimal performance on a given
computer.  In addition, it is important for the community to increase
the capability of the data pipeline applications it has developed (e.g.
IRAF, IDL, ...).  Currently, these tools provide a variety of
application specific functions.  Unfortunately, they are not designed
for real-time parallel data pipelines.  It would be highly beneficial 
to upgrade (and add to) these tools so that they can exploit the
hardware technology that will be required to effectively process large
surveys.

\section{Summary}
  The NVOs core data mining and archive federation activities are
heavily dependent on the underlying data pipeline software necessary to
translate the raw data into scientifically relevant source detections. 
The data pipeline software dictates: the raw data storage and retrieval
mechanisms, the meaning and format of the fields in the source catalogs,
and the ability of the NVO users to re-analyze raw data for their own
purposes.  Increasing the performance of the core data pipeline
software so that it can address the needs of current and future high
data rate surveys is an important activity that should be addressed in
concert with the development of the NVO.

\begin{figure}
%\plotone{pipeline.eps}
\plotone{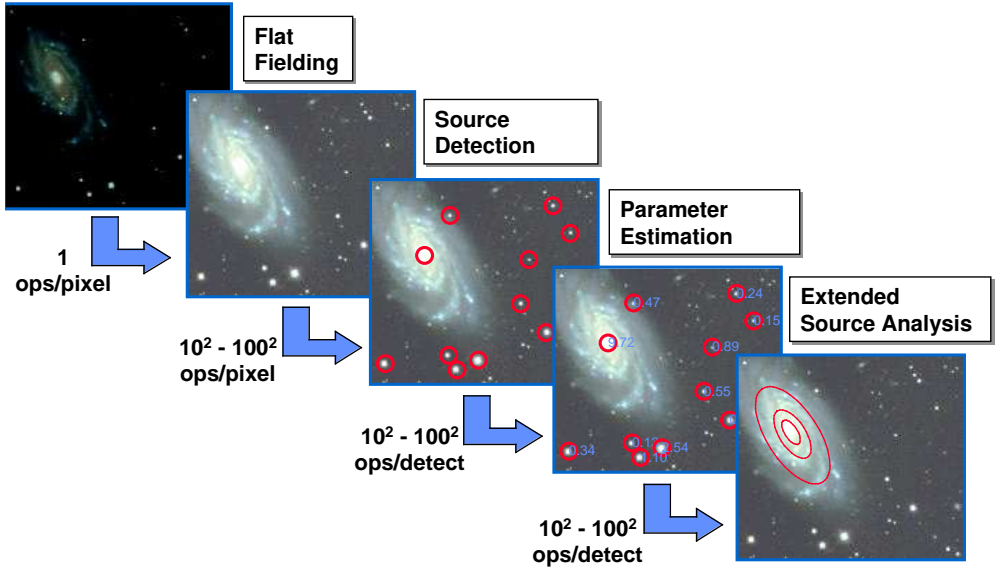}
\caption{
Standard steps in a data reduction pipeline.
The computing requirements to fulfill the operation are shown
below each step.
}
\end{figure}

\begin{figure}
%\plotone{hierarchy.eps}
\plotone{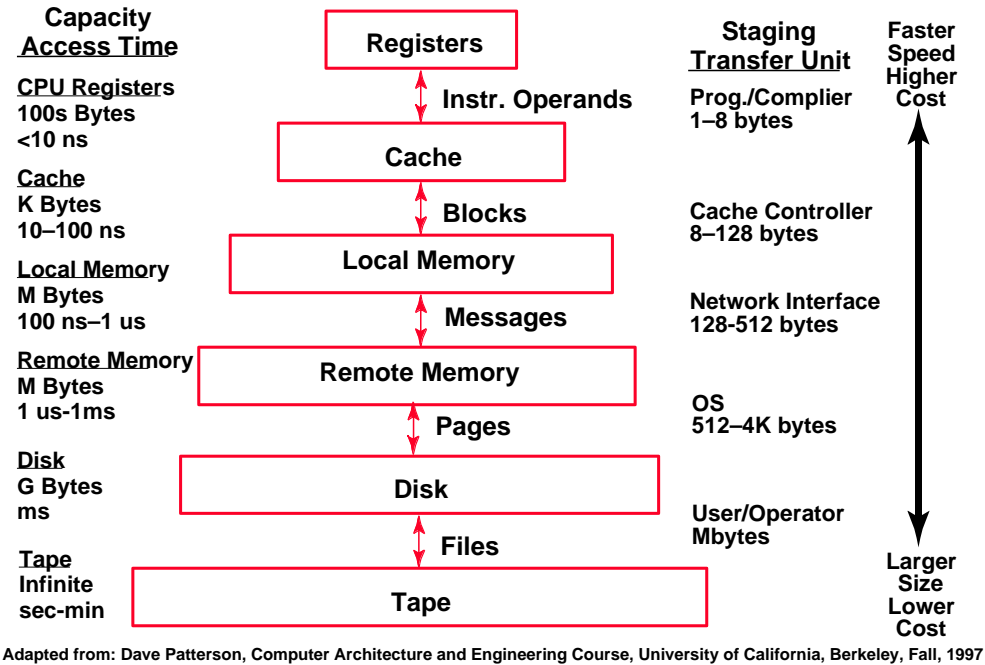}
\caption{
The complex memory hierarchy of a modern computer system.
Typically, a data reduction pipeline incorporates the details
of this memory hierarchy into the program, which limits portability
and re-use of pipeline software.
}
\end{figure}

\begin{figure}
%\plotone{interconnects.eps}
\plotone{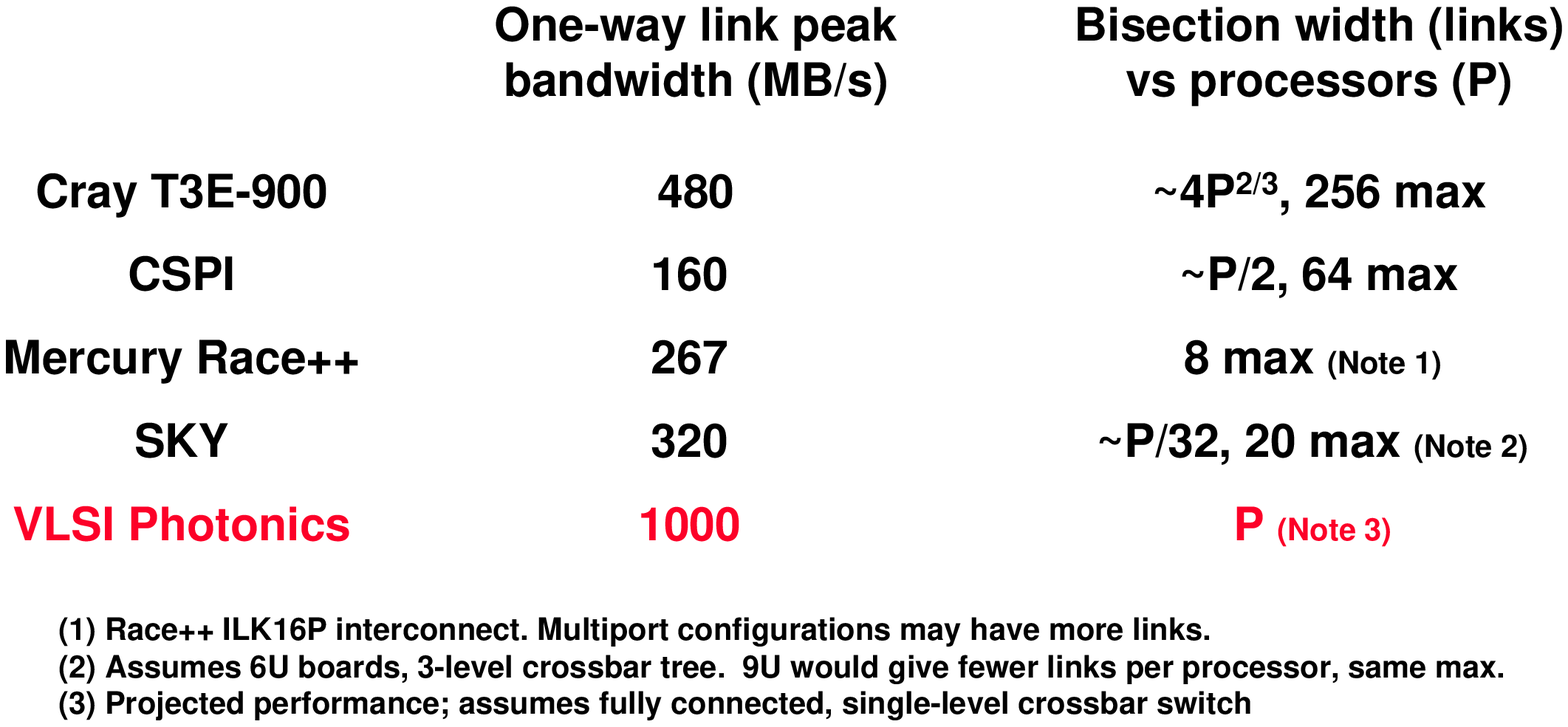}
\caption{
Current and future capabilities of various interconnect
technologies.
}
\end{figure}

\begin{figure}
%\plotone{libraries.eps}
\plotone{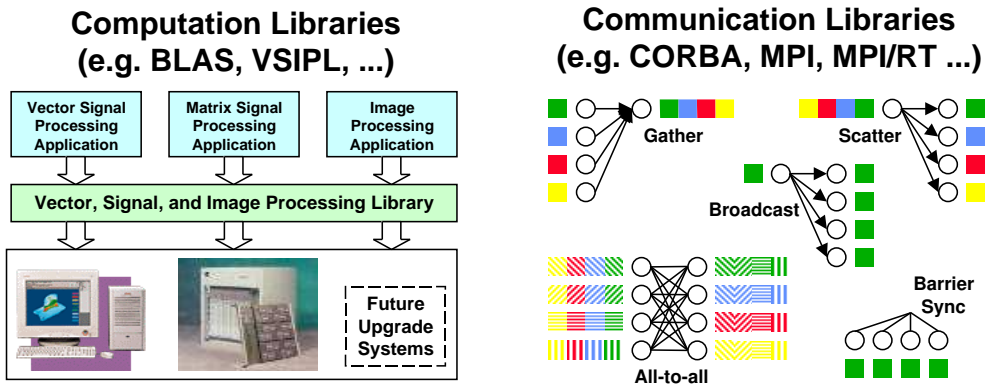}
\caption{
Summary of various existing software libraries and their capabilities.
}
\end{figure}

\end{document}